\documentclass[11pt]{article}
\usepackage{latexsym}
\usepackage{amsmath,amsthm,amstext,amscd,amssymb,euscript,bbm,color}
\usepackage{graphicx}
\newtheorem{theorem}{Theorem}[section]
\newtheorem{definition}[theorem]{Definition}
\newtheorem{remark}[theorem]{Remark}
\newtheorem{example}[theorem]{Example}
\newtheorem{lemma}[theorem]{Lemma}

\newtheorem{proposition}[theorem]{Proposition}
\newtheorem{corollary}[theorem]{Corollary}

\newcommand{\beq}{\begin{equation}}
\newcommand{\eeq}{\end{equation}}

\newcommand{\Zd}{\mathbb Z^d}

\newcommand{\N}{\mathbb N}
\newcommand{\R}{\mathbb R}
\newcommand{\E}{\mathbb E}

\newcommand{\be}{\begin{equation}}
\newcommand{\ee}{\end{equation}}
\newcommand{\bl}{\begin{lemma}}
\newcommand{\el}{\end{lemma}}

\newcommand{\br}{\begin{remark}}
\newcommand{\er}{\end{remark}}

\newcommand{\bt}{\begin{theorem}}
\newcommand{\et}{\end{theorem}}

\newcommand{\bd}{\begin{definition}}
\newcommand{\ed}{\end{definition}}

\newcommand{\bp}{\begin{proposition}}
\newcommand{\ep}{\end{proposition}}

\newcommand{\bc}{\begin{corollary}}
\newcommand{\ec}{\end{corollary}}

\newcommand{\bpr}{\begin{proof}}
\newcommand{\epr}{\end{proof}}

\newcommand{\bi}{\begin{itemize}}
\newcommand{\ei}{\end{itemize}}

\newcommand{\ben}{\begin{enumerate}}
\newcommand{\een}{\end{enumerate}}

\newcommand{\pee}{\ensuremath{\mathbb{P}}}
\newcommand{\ce}{\ensuremath{\mathcal{C}}}

\newcommand{\loc}{\ensuremath{\mathcal{L}}}

\newcommand{\fe}{\ensuremath{\mathcal{F}}}
\newcommand{\mee}{\ensuremath{\mathcal{M}}}

\newcommand{\si}{\ensuremath{\sigma}}

\newcommand{\epsi}{\ensuremath{\epsilon}}

\newcommand{\caD}{{\mathcal D}}

\newcommand{\caK}{{\mathcal K}}

\newcommand{\caT}{{\mathcal T}}

\newcommand{\lij}{ {\left(x_i\frac{\partial}{\partial x_j}-x_j\frac{\partial}{\partial x_i}\right)^2}}
\newcommand{\lija}{{\left(x_{i,\alpha}\frac{\partial}{\partial x_{j,\beta}}-x_{j,\beta}\frac{\partial}{\partial x_{i,\alpha}}\right)^2}}

\begin{document}
\title{{\bf Duality and hidden symmetries in interacting particle systems}}
\author{
Cristian Giardin\`a
\footnote{Department of Mathematics and Computer Science,
Eindhoven University, P.O. Box 513 - 5600 MB Eindhoven, The Netherlands,
{\em c.giardina@tue.nl}}\\
Jorge\ Kurchan
\footnote{CNRS-Ecole Sup\'erieure de Physique et de Chimie Industrielles , rue Vauquelin 10, 75231 Paris, France,
{\em jorge@pmmh.espci.fr}}\\
Frank\ Redig
\footnote{Mathematisch Instituut Universiteit Leiden,
Niels Bohrweg 1, 2333 CA Leiden, The Netherlands,
{\em redig@math.leidenuniv.nl}}\\
Kiamars\ Vafayi
\footnote{Mathematisch Instituut Universiteit Leiden,
Niels Bohrweg 1, 2333 CA Leiden, The Netherlands,
{\em vafayi@math.leidenuniv.nl}}
}
\vspace{2.cm}
\maketitle
\begin{quote}

Abstract: In the context of Markov processes, both in discrete
and continuous setting, we show a general relation between
duality functions and symmetries of the generator.
If the generator can be written in the form of
a Hamiltonian of a quantum spin system, then the ``hidden''
symmetries are easily derived.
We illustrate our approach in processes of symmetric exclusion type,
in which the symmetry is of $SU(2)$ type,
as well as for the Kipnis-Marchioro-Presutti (KMP) model for which
we unveil its $SU(1,1)$ symmetry.
The KMP model is in turn an instantaneous thermalization limit
of the energy process associated to a large family of models of 
interacting diffusions, which we call Brownian energy process (BEP)
and which all possess the $SU(1,1)$ symmetry.
We treat in details the case where the system is in contact with 
reservoirs and the dual process becomes absorbing.

\end{quote}
\vspace{12pt}
\newpage
\section{Introduction}
Duality is a technique developed in the
probabilistic literature that allows to obtain elegant and
general solutions of some problems in interacting particle systems.
One transforms the evaluation of a correlation function in the
original model to a simpler quantity in the dual one.

The basic idea of duality in interacting particle systems goes back
to Spitzer  \cite{Spitzer} who introduced it for
symmetric exclusion process (SEP) and independent random walkers to
characterize the stationary distribution. Later, Ligget
\cite{Liggett} systematically introduced duality for spin systems
and used it, among others, for the complete characterization of
ergodic properties of SEP, voter model, etc. Duality property might
also be useful in the context of transport models and
non-equilibrium statistical mechanics, that is when the bulk
particle systems is in contact at its boundaries with reservoirs
working at different values of their parameters. For instance,
considering again the symmetric exclusion process in contact with
particle reservoirs at different chemical potentials, Spohn used
duality to compute the $2$-point correlation function \cite{S},
showing the existence of long-range correlations in non-equilibrium
systems. In the case of energy transport, i.e. interacting particle
systems with a continuous dynamical variable (the energy) connected
at their boundaries to thermal reservoirs working at different
temperatures, duality has been constructed for the
Kipnis-Marchioro-Presutti (KMP) model \cite{KMP} for heat conduction
and also for other models \cite{gkr}. Consequences of duality
include the possibility to express the $n$-point energy correlation
functions in terms of $n$ (interacting) random walkers. Duality has
also been used in the study of biological population models, see
\cite{mohle} and references therein.

One should notice that the construction of a dual process is usually
performed with an ad-hoc procedure which requires the ansatz of a
proper duality function on which the duality property can be
established. The closure of $n$-point correlations functions at each
order might be an indication that a dual process exists. However in
the general case the closure property is neither sufficient nor
necessary to construct the dual process. In this paper we present a
general procedure to derive a duality function and a dual process
from the symmetries of the original process. When applied to
transport models, our theorems allow to identify the source of the
existence of a dual process with the non-abelian symmetries of the
evolution operator. The idea is simple: transport models have in the
bulk a symmetry associated with a conserved quantity, the one that
is transported. It may happen in some cases that this symmetry is a
subgroup of a larger group, i.e. that extra (less obvious) symmetry
are present. In that case, one can describe the same physical
situation as the transport of another quantity (another element of
the group), and in some cases this makes the problem simpler. In the
physics literature Sandow and Schutz \cite{schutz} realized that
this is case for the SEP process, whose $SU(2)$ symmetry they made
explicit by writing the evolution operator in quantum spin notation.
In this paper we study in full generality the relation between
duality and symmetries. We give a general scheme for constructing
duality for continuous time Markov processes whose generator has a
symmetry. For interacting particle systems used as transport models
we detail the effect of the reservoirs. For particle transport
models we generalize the symmetric exclusion process to a situation
where each site can accommodate up to $2j$ particles, with
$j\in\N/2$. For energy transport models we uncover a hidden
$SU(1,1)$ symmetry in a large class of models for energy transport
(including KMP model) which explains their duality property, as the
$SU(2)$ does for the SEP process.

\section{Definitions and Results}
\label{sez2}

\subsection{Generalities}
Let $(\eta_t)_{t\ge 0}$ denote a Markov process on a state space $\Omega$.
Elements of the state space are denoted by $\eta,\xi,\zeta,$..
The probability measure on path space starting from $\eta$
is called $\pee_{\eta}$, and $\E_{\eta}$ denotes
expectation with respect to $\pee_{\eta}$.
In the whole of this paper, we will restrict to Feller processes.
In that case, to the process $(\eta_t)_{t\ge 0}$
there corresponds a strongly continuous,
positivity-preserving, contraction semigroup
$A_t: {\cal C}(\Omega)\to {\cal C}(\Omega)$
with domain the set ${\cal C}(\Omega)$ of continuous functions $f:\Omega \to \R$
\begin{equation}
\label{aspetta}
A_t f(\eta) := \E_{\eta} f(\eta_t) =
\E(f(\eta_t)|\eta_0 = \eta) = \int f(\eta')p_t(\eta,d\eta')
\end{equation}
where $p_t(\eta,d\eta')$ is the transition kernel of the process.
The infinitesimal generator of the semigroup is denoted by $L$,
$$
Lf = \lim_{t\rightarrow 0} \frac{{A}_t f - f }{t}
$$
and is defined on its natural domain, i.e. the set
of functions $f:\Omega \to \R$ for which the limit in the r.h.s. exists
in the uniform metric.
We also consider the adjoint of the semigroup,
with domain ${\cal M}(\Omega)$ the set of signed
finite Borel measures, ${A}^*_t: {\cal M}(\Omega) \to {\cal M}(\Omega)$,
defined by
$$
<f, {A}_t^* \mu > \;=\; <{A}_t f ,\mu >
$$
where the pairing $<\cdot,\cdot>: {\cal C}(\Omega)\times
{\cal M}(\Omega)\to \R$ is given by
$$
<f, \mu> = \int f d\mu
$$
The processes which appear in our applications
will always be either jump process or diffusions.
\begin{example}
\label{example1}
In the case that the Markov process $(\eta_t)_{t\ge
0}$ is a pure jump process and the state space $\Omega$ is finite or
countable then the generator is of the form
$$
{L} f(\eta) = \sum_{\eta'\in\Omega} c(\eta,\eta')(f(\eta') - f(\eta))
$$
where $c(\eta,\eta')\ge 0$ is the rate for a transition from
configuration $\eta$ to configuration $\eta'$. Equivalently
we can write
$$
{L} f(\eta) = \sum_{\eta'\in\Omega} L(\eta,\eta')f(\eta')
$$
where $L(\eta,\eta')$ is a matrix having positive off-diagonal
elements and rows sum equal to zero, namely
$$
L(\eta,\eta') = \left\{
\begin{array}{rl}
c(\eta,\eta') & \text{if }\;\; \eta\ne\eta'\\
-\sum_{\eta''\ne\eta} c(\eta,\eta'') & \text{if }\;\; \eta=\eta'\\
\end{array} \right.
$$
In the context of a countable state space $\Omega$
we have the usual exponential of a matrix, so that
$$
A_t = e^{tL} = \sum_{i=0}^{\infty} (tL)^i/i!
$$
and
$A_t^* = A_t^{T}$ where the superscript $^T$ denotes transposition.
\end{example}
\begin{example}
General diffusion processes with state space $\Omega= \R^N$
are also considered here. In this case the generator take
the form of a differential operator of the second order
$$
{L} f = \sum_{i,j=1}^N a(x_i,x_j) \frac{\partial^2 f}{\partial x_i \partial x_j}
+ \sum_{i=1}^N b(x_i) \frac{\partial f}{\partial x_i}
$$
(see \cite{SV} for general conditions which guarantees
that $L$ satisfy the maximum principle and thus generate a
positivity preserving semigroup).
\end{example}
\subsection{Duality and Self-duality}

\begin{definition}[self-duality]
Consider two independent copies $(\eta_t)_{t\ge 0}$ and $(\xi_t)_{t\ge 0}$
of a continuous time Markov processes on a state space $\Omega$.
We say that the process is self-dual with self-duality function
$D:\Omega\times\Omega\to \R$ if for all $(\eta,\xi)\in\Omega\times\Omega$,
we have
\be
\label{selfdualdef}
\E_\eta D(\eta_t,\xi)= \E_{\xi} D(\eta,\xi_t)
\ee
\end{definition}
\begin{definition}[duality]
\label{dualllity}
Consider two continuous time Markov processes: $(\eta_t)_{t\ge 0}$
on a state space $\Omega$ and $(\xi_t)_{t\ge 0}$
on a state space $\Omega_{dual}$.
We say that $(\xi_t)_{t\ge 0}$ is the dual of $(\eta_t)_{t\ge 0}$
with duality function $D:\Omega\times\Omega_{dual}\to \R$ if
for all $\eta\in\Omega$, $\xi\in\Omega_{dual}$ we have
\be
\label{dualdef}
\E_\eta D(\eta_t,\xi)= \E^{dual}_{\xi} D(\eta,\xi_t)
\ee
\end{definition}
If $A_t$ denotes the semigroup of the original process
$(\eta_t)_{t\ge 0}$  and  $A_t^{dual}$ denotes the
semigroup of the related dual process $(\xi_t)_{t\ge 0}$ then,
using Eq. \eqref{aspetta},
the definition \ref{dualllity} is equivalent to
\be
\label{dualityandgen}
A_t D (\eta,\xi)= A_t^{dual} D (\eta,\xi)
\ee
where it is understood that on the l.h.s.
of \eqref{dualityandgen} the operator $A_t$ works on the $\eta$ variable,
while on the r.h.s. the operator$A_t^{dual}$
works on the $\xi$ variable.

If the original process
$(\eta_t)_{t\ge 0}$  and  the dual process
$(\xi_t)_{t\ge 0}$  are Markov processes with finite or
countably infinite state space $\Omega$, resp. $\Omega_{dual}$,
(cfr. Example \ref{example1}) property \eqref{dualityandgen}
is equivalent with its ``infinitesimal version''
in terms of the generators
\be \label{gendualdef}
\sum_{\eta'\in\Omega}L(\eta,\eta')D(\eta',\xi)= \sum_{\xi'\in\Omega}
L_{dual}(\xi,\xi') D(\eta,\xi')
\ee
In matrix notation, this reads
\be
\label{dualmagen}
LD = DL_{dual}^T
\ee
where $D$ is the matrix with elements
$D(\eta,\xi)$ and $(\eta,\xi)\in\Omega\times\Omega_{dual}$.
Remark that in this case $D$ is not necessarily a square matrix, because the
state spaces $\Omega$ and $\Omega_{dual}$ are not necessarily equal
and or of equal cardinality.

When $\Omega = \Omega_{dual}$ and $A_t=A_t^{dual}$,
then an equivalent condition for self-duality (cfr. (\ref{selfdualdef})) is
\be
\label{dualma}
LD = DL^T
\ee

\subsection{Duality and Symmetries}
We first discuss self-duality and then duality.
We consider the simple context of finite or countably infinite state space Markov processes.
In many cases of interacting particle systems, the generator is a sum of operators
working only on a finite set of coordinates of the configuration.
Therefore, showing (self)-duality reduces to showing (self)-duality for the
individual terms appearing in this sum, which is a finite state
space situation.

\begin{definition}
Let $A$ and $B$ be two matrices having the same dimension. We say that $A$ is a symmetry of $B$
if $A$ commutes with $B$, i.e.
\be\label{symdef}
AB=BA
\ee
\end{definition}

The first theorem shows that self-duality functions and symmetries are in
one-to-one correspondence, provided $L$ and $L^T$ are similar
matrices, which is automatically the case in the finite state space
context.

\bt
\label{king}
Let $L$ be the generator of a finite or countable
state space Markov process. Let $Q$ be a matrix such that
\be\label{conj} L^T = Q L Q^{-1} \,.\ee
Then we have
\ben
\item If $S$ is a symmetry of the generator,
then $SQ^{-1}$ is a self-duality function.
\item If $D$ is a self-duality function, then
$DQ$ is a symmetry of the generator.
\item If $S$ is a symmetry of $L^T$, then $Q^{-1}S$
is a self-duality function
\item If $D$ is a self-duality function,
then $QD$ commutes with $L^T$.

\een \et
\begin{proof}
The proof is elementary. We show items 1 and 2
(item 3 and 4 are obtained in a similar manner).
Combining \eqref{conj} with
\eqref{symdef}, we find \be\label{dualsym} L(SQ^{-1}) = (S Q^{-1})
L^T \ee i.e., $D=SQ^{-1}$ is a self-duality function (see Eq. \eqref{dualma}).
Conversely, if $D$ is a self-duality function, then
combining \eqref{conj} with \eqref{dualma} one proves
\eqref{symdef} for $S =DQ$.
\end{proof}

\begin{remark}
Self-duality functions are not unique, i.e. there might exist
several self-duality functions for a process. This is evident form
the fact that if D is a duality function for self-duality, and $S$
is a symmetry, then $SD$ is also a duality function for
self-duality. An interesting question is to study the vector space
of self-duality functions, its dimension, etc. However this
question is not addressed in this paper. See \cite{mohle} for a discussion
of this issue and some examples
in the context of Markov processes with discrete state space.
\end{remark}

\begin{remark}
\label{simildet}
In the finite state space context, $L$ and $L^T$ are
always similar matrices \cite{TZ}, i.e., there exists a conjugation
matrix $Q$ such that $L^T = Q L Q^{-1}$.
In interacting particle system the matrix $Q$ can usually be easily
constructed. As an example, in the case that ${\cal L}$ has a reversible
measure, i.e., a probability measure $\mu$ on $\Omega$ such that
\be\label{detbal} \mu(\eta)L(\eta,\eta')=\mu(\eta')L(\eta',\eta) \ee
for all $\eta,\eta'\in\Omega$, then a diagonal conjugation
matrix $Q$ is given by \be\label{invconj} Q(\eta,\eta')=
\mu(\eta) \delta_{\eta,\eta'} \ee In general, if $\mu$ is a
stationary measure then
$$
L_{rev}(\eta,\xi):= \frac{L(\xi,\eta) \mu(\eta)}{\mu(\xi)}
$$
is the generator of the time-reversed process, which is clearly
similar to $L^T$. Therefore, the similarity of $L$ and $L^T$ is
equivalent with the similarity of the generator and the
time-reversed generator.
\end{remark}

Self-duality is a particular case of duality.
To generalize Theorem \ref{king} to the context of
(general) duality we need the notion
of conjugation between two matrices.
\begin{definition}
\label{conjugate}
Let $A$ be a matrix of dimension $m\times m$ and
let $B$ be a matrix of dimension $n\times n$.
$A$ and $B$ are called {\em conjugate} if there
exist matrices $C$ of dimension $m\times n$ and
$\tilde{C}$ of dimension $n\times m$ such that
\be
AC = CB, \qquad \tilde{C} A = B \tilde{C}
\ee
\end{definition}
We then have the following analogue of Theorem \ref{king}.

\bt
\label{tsar}
Let $L$ and $L_{dual}$ be generators of finite or countable state
space Markov chains. Then we have the following.
\ben
\item
If $Q$ is the matrix that gives the similarity
\be\label{simil2}
L_{dual}^T = QL_{dual} Q^{-1}
\ee
and $C$ and $\tilde C$ are the matrices giving
the conjugacy between $L$ and $L_{dual}$ in the sense
of definition \ref{conjugate}, then:
\ben
\item For any symmetry $S$ of the generator $L$, $D=SCQ^{-1}$ is a duality function.
\item If $D$ is a duality function, then $S= DQ\tilde{C}$ is a symmetry of $L$.
\een
\item
If $Q$ is the matrix that gives the similarity
\be\label{simil3}
L^T = QL Q^{-1}
\ee
and $C$ and $\tilde C$ are the matrices giving
the conjugacy between $L^T$ and $L_{dual}^T$ in the sense
of definition \ref{conjugate}, then:
\ben
\item For any symmetry $S$ of the transposed generator $L^T$, $Q^{-1}SC$ is a duality function.
\item If $D$ is a duality function, then $QD\tilde{C}$ commutes with $L^T$.
\een
\een
\et

\begin{proof}
The proof of item 1(a) is given by the following series of equalities
\be\label{gensymdualma}
 L(SCQ^{-1}) = SLC Q^{-1}= SCL_{dual} Q^{-1}= (SCQ^{-1}) L_{dual}^T
\ee
The first equality uses the hypothesis of $S$ being a symmetry
of the generator $L$,
the second comes from the conjugation of the generators, the third
is obtained from the similarity transformation \eqref{simil2}.
If one recall \eqref{dualmagen} then Eq.\eqref{gensymdualma}
shows that $D= SCQ^{-1}$ is a duality function.
The proof of the other items follow from a similar argument.
\end{proof}

\section{Examples with two sites}

In this section we present a series of examples where particles
jump on two lattice sites. We wish to show how (self)-duality
can be established by making use of the previous theorems.
To identify the symmetries we will rewrite the stochastic
generator, or its adjoint, in terms of generators of some
symmetry group. Some of the examples will be useful later for
the study of transport models. In fact, many transport models
such as the exclusion process have a generator that is
written as the sum of operators working on two sites.

\subsection{Self-duality for symmetric exclusion}
We first recover the
classical self-duality for symmetric exclusion \cite{Liggett}.
One has two sites (labeled $1,2$) and configurations
have at most one particle at each site.
Particles hop at rate one from one site to another, and jumps
leading to more than one particle at a site are suppressed.
As usual we write $0,1$ for absence resp.\ presence of particle.
The state space is then $\Omega = \{00,01,10,11\}$. Elements
in the state space are denoted as $\eta = (\eta_1\eta_2)$.
The matrix elements of the generator are given by
$L_{01,10}=L_{10,01}=1=-L_{01,01}=-L_{10,10}$, and all other
elements are zero.

To apply Theorem \ref{king} we need to identify a symmetry $S$
of the generator.
The transposed of the generator can be written
as
\be\label{sep12gen}
L^T= J_1^+ \otimes J_2^{-} + J_1^- \otimes J_2^+ + 2J^0_1 \otimes J^0_2 - \frac12 \mathbbm{1}_1 \otimes  \mathbbm{1}_2
\ee
where the operators $J_i^a$ with $i\in\{1,2\}$ and $a\in\{+,-,0\}$ act on a $2$-dimensional Hilbert space,
with basis $|0\rangle = {1\choose 0}, |1\rangle = {0\choose 1}$, as
\be\label{jee}
J_i^+=\left(
\begin{array}{cc}
0 &0\\
1 &0\\
\end{array}
\right)
\qquad
J_i^-=\left(
\begin{array}{cc}
0 &1\\
0 &0\\
\end{array}
\right)\qquad
J_i^0=\left(
\begin{array}{cc}
-1/2 &0\\
0 & 1/2\\
\end{array}
\right)
\ee
and $\mathbbm{1}_i$ is the identity matrix.
The operators $J_i^a$ with $a\in\{+,-,0\}$ satisfy the $SU(2)$ commutation relations:
\begin{eqnarray}\label{su2com}
[J_i^0, J_i^{\pm}] &=& J_i^{\pm}
\nonumber\\
{[} J_i^-, J_i^+ ] &=& -2J_i^0
\end{eqnarray}
from which we deduce (cfr \eqref{sep12gen}) that
$L^T$ commutes with the three generators of the $SU(2)$ group,
$J^{a}= J_1^{a}\otimes\mathbbm{1}_2 + \mathbbm{1}_1\otimes J_2^{a}$ for $a\in\{ +,-,0\}$.
A possible choice for the symmetry of $L^T$ is then obtained by
considering the creation operator $J^+$ and exponentiating in order
to have a factorized form
\[
 S = e^{J^{+}} = e^{J_1^+\otimes \mathbbm{1}_2 + \mathbbm{1}_1 \otimes  J_2^+} =
 e^{J_1^+} \otimes e^{J_2^+} = S_1 \otimes S_2
\]
More explicitly, in the basis $|0\rangle\otimes|0 \rangle,|0\rangle\otimes|1\rangle,|1\rangle\otimes|0\rangle,
|1\rangle\otimes|1\rangle,$ the matrix $S$ is
\[
S=\left(
\begin{array}{cc}
1 &0\\
1 &1\\
\end{array}
\right)
\otimes
\left(
\begin{array}{cc}
1 &0\\
1 &1\\
\end{array}
\right)
=
\left(
\begin{array}{cccc}
1 &0&0 &0\\
1 &1&0 &0\\
1&0&1&0\\
1&1&1&1\\
\end{array}
\right)
\]
We also need the similarity transformation between $L$ and $L^T$.
The matrix $Q$, relating $L$ to its transposed, is the identity
since $L$ is symmetric. A duality function for self-duality is
thus given by $D= Q^{-1}S = S$. Notice that $D$ can also be written
as
$$
D(\eta_1\eta_2,\xi_1\xi_2) = \prod_{i\in\{1,2\}:\xi_i = 1} \eta_i\,
$$
which is the usual self-duality function of \cite{Liggett}.

\subsection{Self-duality for $2j$-symmetric exclusion}

Now we consider two sites with at most $2j$ particles
on each site, with $j\in\N/2$. The state space is
$\Omega = \Omega_1\times\Omega_2$ where $\Omega_i = \{0,1,\ldots 2j\}$.
The rates for transitions are the following:
if there are $\eta_1$ particles at site
$1$ and $\eta_2$ particles at site $2$, a particle
is moved from $1$ to $2$ at rate $\eta_1 (2j-\eta_2)$ and
from $2$ to $1$ at rate $\eta_2 (2j-\eta_1)$.
So in this case the generator is given by
\begin{eqnarray*}
L(\eta_1\eta_2 ,\eta_1'\eta_2') & = &
\eta_1 (2j-\eta_2)\delta_{\eta_1-1, \eta_1'}\delta_{\eta_2+1, \eta_2'} +
\eta_2 (2j-\eta_1)\delta_{\eta_1+1, \eta_1'}\delta_{\eta_2-1,\eta_2'}
\\
& &
- (\eta_1(2j-\eta_2)+\eta_2 (2j-\eta_1))\delta_{\eta_1,\eta_1'}\delta_{\eta_2,\eta_2'}
\end{eqnarray*}
The transposed of this generator can also be expressed as
the scalar product between two spin operators satisfying the
$SU(2)$ algebra, namely
\be\label{sepjgen}
L^T= J_1^+ \otimes J_2^{-} + J_1^- \otimes J_2^+ + 2J^0_1\otimes J^0_2 -2j^2 \mathbbm{1}_1 \otimes  \mathbbm{1}_2
\ee
where the $J_i^a$, $i\in\{1,2\}$ and $a\in\{+,-,0\}$, act on a $(2j+1)$-dimensional Hilbert space
with orthonormal basis $|0\rangle, |1\rangle, \ldots |2j\rangle$ as
\begin{eqnarray}\label{su2op}
J_i^+|\eta_i\rangle &=& (2j-\eta_i) |\eta_i+1\rangle\nonumber\\
J_i^-|\eta_i\rangle &=& \eta_i|\eta_i-1\rangle\nonumber\\
J_i^0|\eta_i\rangle &=& (\eta_i-j) |\eta_i\rangle
\end{eqnarray}
The standard symmetric exclusion process of the previous section
is recovered when $j=1/2$.
Reasoning as above, a symmetry of the generator is
\[
S = S_1 \otimes S_2 = e^{J_1^+} \otimes e^{J_2^+}
\]
which has matrix elements $S(\eta_1\eta_2, \xi_1\xi_2) = S_1(\eta_1,\xi_1)S_2(\eta_2,\xi_2)$ with
\be
\label{cc1}
S_i(\eta_i,\xi_i) = \langle\eta_i|e^{J_i^+}|\xi_i\rangle = { {2j-\eta_i} \choose {\eta_i-\xi_i}}
\ee
where we adopt the convention ${n\choose m}=0$ for $m>n$.

To detect the matrix $Q$ giving the similarity transform between $L$ and $L^T$
(notice that $L$ is not symmetric anymore for $j\ne 1/2$) we make use of remark
\ref{simildet} and use the fact that the invariant measures of the
$2j$-symmetric exclusion process are products of binomials $Bin(2j, \rho)$,
with a free parameter $0<\rho<1$ (this will be proved in Theorem \ref{sepjthm}).
Therefore, if we choose $\rho=1/2$ then a possible choise is $Q = Q_1\otimes Q_2$ with
\be
\label{cc2}
Q_i(\eta_i,\eta_i') = \delta_{\eta_i,\eta_i'} {2j\choose \eta_i}
\ee
Combining \eqref{cc1} and \eqref{cc2}, Theorem \ref{king} then implies
that a duality function for self-duality is given by
$$
D = D_1 \otimes D_2 = Q_1^{-1} S_1 \otimes Q_2^{-1} S_2
$$
with
\be
\label{dual2jfunction}
D_i(\eta_i,\xi_i )= (Q_i^{-1} S_i) (\eta_i,\xi_i ) = \frac{{\eta_i\choose \xi_i}}{{2J\choose \xi_i}}
\ee
Later, in Theorem \ref{sepjthm}, we will give a probabilistic interpretation
of this function.

\subsection{Self-duality for the dual-BEP}

This is a process that can be viewed as a ``bosonic'' analogue
of the SEP (particles attract each other rather than repel with
the exclusion hard core constraint). The state space
is $\Omega = \Omega_1\times\Omega_2$ with $\Omega_i = \N$,
i.e. we have two sites each of which can accommodate an unlimited
number of particles. For $\eta_1$ particles
at site $1$, $\eta_2$ particles at site $2$, the
rate of putting a particle from $1$ to $2$
is given by $2\eta_1(2\eta_2+1)$ and the rate of
moving a particle from $2$ to $1$
is given by $2\eta_2(2\eta_1+1)$ .
We will see later how this process arises naturally
as a dual of the Brownian Energy Process (BEP),
see Section \ref{secmep} below.

The matrix of the generator is given by
\begin{eqnarray}
 L(\eta_1\eta_2,\eta_1' \eta_2')
 & = &
 2\eta_1(2\eta_2+1) \delta_{\eta_1',\eta_1-1}\delta_{\eta_2',\eta_2+1}
 + 2\eta_2(2\eta_1+1)\delta_{\eta_1',\eta_1+1}\delta_{\eta_2',\eta_2-1}
\nonumber\\
& & -(8\eta_1\eta_2+2\eta_1+2\eta_2)\delta_{\eta_1,\eta_1'}\delta_{\eta_2,\eta_2'}.
\end{eqnarray}
The transposed of the generator can be written in terms of
generators of a $SU(1,1)$ algebra as follows.
On each site $i\in\{1,2\}$ we consider
operators $K_i^{a}$ with $a\in \{ +,-, 0\}$ given by
\begin{eqnarray}\label{su11op}
K_i^+ |\eta_i\rangle &=& (\eta_i+ 1/2) |\eta_i+1\rangle\nonumber\\
K_i^- |\eta_i\rangle &= & \eta_i|\eta_i-1\rangle\nonumber\\
K_i^0 |\eta_i\rangle &=& (\eta_i+ 1/4) |\eta_i\rangle
\end{eqnarray}
They satisfy the commutation relations of $SU(1,1)$:
\begin{eqnarray}\label{su11com}
[K_i^0, K_i^\pm] &=& \pm K_i^\pm \nonumber\\
{[} K_i^-, K_i^+] &= & 2 K_i^0
\end{eqnarray}
The transposed of the generator then reads
\be\label{su11gen}
L^T= 4\left(K_1^+ \otimes K_2^- + K_1^- \otimes K_2^+ - 2K^0_1\otimes K^0_2 + \frac{1}{8}\mathbbm{1}_1 \otimes  \mathbbm{1}_2\right)
\ee
From the commutation relations, it is easy to see that
$L^T$ commutes with $K^a = K_1^a\otimes \mathbbm{1}_2 + \mathbbm{1}_1 \otimes K_2^a$,
for $a\in \{ +, -, 0\}$.
A possible symmetry is then given by the matrix
\[
S = S_1 \otimes S_2 = e^{K_1^+}\otimes e^{K_2^+}
\]
which has matrix elements $S(\eta_1\eta_2, \xi_1\xi_2) = S_1(\eta_1,\xi_1)S_2(\eta_2,\xi_2)$ with
\be
\label{dd1}
S_i(\eta_i,\xi_i) = \langle \eta_i| e^{K_1^+} | \xi_i\rangle = \frac{(2\eta_i-1)!!}{(2\xi_i-1)!! (\eta_i-\xi_i)! \;2^{\eta_i -\xi_i}}
\ee
A similarity transformation $L^T = Q^{-1} L Q$ to pass to the transposed
is suggested (remark \ref{simildet}) by the knowledge of the stationary measure
of the dual-BEP (see Theorem \ref{bosmes})
\[
 Q(\eta_1\eta_2,\eta_1' \eta_2') = Q_1(\eta_1,\eta_1')Q_2(\eta_2,\eta_2')
\]
with
\be
\label{dd2}
 Q_i(\eta_i,\xi_i) = \delta_{\eta_i,\xi_i} \left(\frac{\eta_i!}{(2\eta_i-1)!!} 2^{\eta_i}\right)^{-1}
\ee
The self-duality function corresponding to $S$ of \eqref{dd1} and $Q$ of \eqref{dd2}
then
reads
$$
D(\eta_1\eta_2,\xi_1 \xi_2) = D_1(\eta_1,\xi_1)D_2(\eta_2,\xi_2)
$$
\be\label{dualfusu11}
D_i(\eta_i,\xi_i) = Q^{-1} (\eta_i,\eta_i)S_i(\eta_i,\xi_i) = 2^{\xi_i}\frac{\eta_i!}{(\eta_i-\xi_i)! (2\xi_i-1)!!}
\ee

\subsection{Self-duality for independent random walkers}

This is a classical example which is included here
for the sake of completeness. We have two site $1$ and $2$, and
particles hop independently from $1$ to $2$ and
from $2$ to $1$ at rate one.
So the rate to put a particle from $1$ to $2$
in a configuration with $\eta_1$ particles at $1$ and
$\eta_2$ particles at $2$ is simply $\eta_1$.
The generator is given by the matrix
\[
 L(\eta_1\eta_2,\eta_1' \eta_2') = \eta_2\delta_{\eta_1,\eta_1'-1}\delta_{\eta_2, \eta_2'+1} + \eta_1\delta_{\eta_1,\eta_1'+1}\delta_{\eta_2,\eta_2'-1}
+ (-\eta_1-\eta_2) \delta_{\eta_1,\eta_1'}\delta_{\eta_2,\eta_2'}
\]
A self-duality function is $D = D_1\otimes D_2$ with
\[
 D(\eta_i,\xi_i) = \frac{\eta_i!}{(\eta_i-\xi_i)!}
\]
The invariant measures are product of Poisson distributions and a possible conjugation is thus given by
$Q = Q_1\otimes Q_2$ with
\[
Q_i (\eta_i,\xi_i) = \delta_{\eta_i,\xi_i} \frac1{\eta_i!}
\]
As a consequence, a symmetry of the generator is given by
$S = S_1\otimes S_2$ with
\be
\label{ppp}
S_i(\eta_i,\xi_i) = (D_iQ_i)(\eta_i,\xi_i) =\frac1{(\eta_i-\xi_i)!}
\ee
This symmetry comes once more from an underlying structure of creation and
annihilation operators satisfying the Heisenberg algebra.
Indeed, if one defines for $i\in\{1,2\}$ operators $a_i^+$ and
$a_i^-$ which are represented on an Hilbert space with basis
$|0\rangle, |1\rangle, |2\rangle, \ldots$
by operators working as
\begin{eqnarray}
\label{inddd}
a_i^{+} |\eta_i\rangle  & = &  |\eta_i+1\rangle
\nonumber\\
a_i^{-} |\eta_i\rangle & = &\eta_i|\eta_i-1\rangle
\end{eqnarray}
then one easily verifies the commutation relation
$$
[a_i^-,a_i^+]=\mathbbm{1}_i\;.
$$ In terms of these matrices,
the transposed of the generator reads \be \label{tiw} L^T = -
(a_1^{+}\otimes\mathbbm{1}_2 -\mathbbm{1}_1\otimes a_2^{+})
(a_1^{-}\otimes\mathbbm{1}_2 -\mathbbm{1}_1\otimes a_2^{-}) \ee
which commutes with $a^{+} = a^+_1\otimes \mathbbm{1}_2 +
\mathbbm{1}_1 \otimes a^+_2$. The symmetry $S$ in \eqref{ppp} is
then recognized as $S=\exp (a_1^+)\otimes \exp(a_2^+)$.

\subsection{Duality between independent random walkers and a deterministic system}
\label{dualdet} As an example of application of
Theorem \ref{tsar} we consider again a system of independent random
walkers jumping between sites $1$ and $2$. We show that this
system is dual to a deterministic system evolving according to
ordinary differential equations.

We consider the ``abstract'' operator ${\cal L}$ \be \label{abstr}
{\cal L} = -(a_1^{+}-a_2^{+})(a_1^{-}-a_2^{-}) \ee where
$a_i^+,a_j^{-}$ are operators satisfying the canonical commutation
relations \be\label{commheis} [a_i^-,a_j^{+}]=\delta_{i,j}
\mathbbm{1}\,. \ee One way to represent the previous operator is by
considering
\[
{a}_{i}^{-} =\frac{\partial}{\partial x_i}, \qquad {a}^{+}_{i} =
x_i, \qquad i\in\{1,2\}
\]
which obviously satisfy  \eqref{commheis}. In this case the operator
\eqref{abstr} takes the form
\[
L = - (x_1-x_2) \left(\frac{\partial}{\partial
x_1}-\frac{\partial}{\partial x_2}\right)
\]
which is the generator of the deterministic system of differential
equations
\begin{eqnarray}\label{rot}
\frac{dx_1 (t)}{dt} &=& -(x_1(t)-x_2(t))
\nonumber\\
\frac{dx_2 (t)}{dt} &=& (x_1(t)-x_2(t))
\end{eqnarray}
with solutions
\begin{eqnarray}\label{sol}
x_1 (t) &=& \frac{x_1(0)+x_2(0)}{2} + \frac{x_1(0)-x_2(0)}{2}
e^{-2t}
\nonumber\\
x_2 (t) &=& \frac{x_1(0)+x_2(0)}{2} - \frac{x_1(0)-x_2(0)}{2}
e^{-2t}
\end{eqnarray}
Another possible way to represent the operator \eqref{abstr} has
just been seen in the previous paragraph. In this case the creation
and annihilation operators are represented as matrices with elements
given by \eqref{inddd} and then the operator \eqref{abstr} can be
seen as the transposed of the generator for a system on independent
random walkers $$ L_{dual}^T = - (a_1^{+}\otimes\mathbbm{1}_2
-\mathbbm{1}_1\otimes a_2^{+}) \circ (a_1^{-}\otimes\mathbbm{1}_2
-\mathbbm{1}_1\otimes a_2^{-})\;.$$ It is immediately checked that
the function
$$
D(x,\xi) = D(x_1,\xi_1) D(x_2,\xi_2)
$$
with
$$
D(x_i,\xi_i) = x_i^{\xi_i}
$$
gives a conjugation between $L$ and $L_{dual}^T$, namely
$$
LD (x,\xi) = DL_{dual}^T (x,\xi)
$$
In this case, this relation reads more explicitely,
\begin{eqnarray}
 && -(x_1-x_2) \left(\frac{\partial}{\partial
x_1}-\frac{\partial}{\partial x_2}\right) (x_1^{n_1} x_2^{n_2})
\nonumber\\
&=&
n_1 x_1^{n_1-1}x_2^{n_2+1} + n_2 x_1^{n_1+1} x_2^{n_2-1} - (n_1+n_2) x_1^{n_1} x_2^{n_2}
\end{eqnarray}

This implies that $x(t)$ and $\xi(t)$ are each other dual with
duality function $D$  and the following relation holds
\be\label{dualblub} x_1(t)^{\xi_1} x_2 (t)^{\xi_2} =
\E_{\xi_1,\xi_2} (x_1 (0)^{\xi_1(t)} x_2 (0)^{\xi_2(t)}) \ee where
the expectation in the rhs is over the independent random walkers
starting from initial configuration with $\eta_1$ particles at $1$
and $\eta_2$ particles at $2$. We will come back to this example in
section \ref{kinf}.

\br
In the last example, we can still use other
representations of the operators $a^-_i,a_i^+$,
satisfying the commutation relation $[a^-_i,a_j^+]= \delta_{ij}$, such that
the abstract operator \eqref{abstr} is the generator of a
Markovian diffusion process.
E.g., if we choose
\begin{eqnarray}\label{booom}
a_i^+ &=& -\frac{\partial}{\partial x_i} + x_i
\nonumber\\
a_i^- &=& \frac{\partial}{\partial x_i}
\end{eqnarray}
then the abstract operator \eqref{abstr} reads
\[
 \loc = \left(\frac{\partial}{\partial
x_1}-\frac{\partial}{\partial x_2}\right)^2 + (x_1-x_2)\left(\frac{\partial}{\partial
x_1}-\frac{\partial}{\partial x_2}\right)
\]
which is the generator of the (degenerate) diffusion:
the ``coordinate $(x_1-x_2)/2$ undergoes a Brownian
motion and $(x_1+x_2)/2$ remains constant.
So in that case we also have duality
\[
 \E_{x_1, x_2} ( D(x_1(t), n_1) D(x_2 (t), n_2) ) = \E_{n_1, n_2} ( D(x_1, n_1(t)) D(x_2 , n_2(t)))
\]
where $D(x,n)$ can be found by the recursion
\[
 D(x,n+1) = a^+ D(x,n)
\]
e.g.\ the first five polynomials are
\[
 D(x,0) =1 , D(x,1) = x, D(x,2) = x^2-1, D(x,3) = -3x + x^3, D(x,4)= 3-6x^2 + x^4
\]

\er
\section{Symmetric exclusion processes}
In this section we study the $2j$-SEP (with $j\in\N/2$),
i.e. exclusion processes with at most $2j$ particles per site,
on a graph $S$. We show how we can understand
self-duality for the $2j$-SEP from ``classical duality''
(in the sense of \cite{Liggett}) of the symmetric exclusion
process on special graphs. We also consider two
limits $j\to\infty$ leading to a deterministic process
or a system of independent random walkers. Finally,
we consider the boundary driven case, and show
that we have a dual with absorbing boundaries.

\subsection{Symmetric exclusion on ladder graphs}
Consider a countable set ${\cal S}$, to be thought of
as the underlying lattice, and a finite set ${\cal I}$ with cardinality
$2j\in\N$. The set ${\cal I}$ is to be thought of as a ``ladder''
on each site with $2j$ levels.

The state space of SEP on the ladder graph ${\cal S}\times {\cal I}$
is $\Omega= \{0,1\}^{{\cal S}\times {\cal I}}$.
A configuration $\zeta\in\Omega$ is called finite if it contains a finite
number of particles, i.e., if $\sum_{i\in {\cal S},\alpha\in {\cal I}} \zeta (i,\alpha) <\infty$.
The process is described as follows. Let $p(i,l)$ denote a symmetric random walk kernel
on ${\cal S}$, i.e., $p(i,l)=p(l,i)\geq 0$, $\sum_{l\in {\cal S}} p(i,l)=1$.
At each site $i\in {\cal S}$ and level $\alpha\in {\cal I}$, there is at most one particle.
Each particle attempts to jump at rate $p(i,l)$ to a site $l\in {\cal S}$ and level $\beta\in {\cal I}$.

More formally, the SEP on a ladder graph ${\cal S}\times {\cal I}$  is the process with the following generator
on local
functions $f:\Omega\to \R$
\be\label{laddersep}
L f(\zeta) = \sum_{i,l \in {\cal S}}\sum_{\alpha, \beta\in {\cal I}} p(i,l) (\zeta(i,\alpha)(1-\zeta(l,\beta))
(f(\zeta^{(i,\alpha),(l,\beta)})- f(\zeta))
\ee
where $\zeta^{(i,\alpha),(l,\beta)}$ denotes the configuration obtained from $\eta$ by
removing a particle at site $i$ level $\alpha$ and placing it at site $l$ level $\beta$.
Since this process is a symmetric exclusion process
on a special graph, then it is self-dual in the following sense:
\bp
\label{old}
Define for $\zeta,\tilde\zeta\in\Omega$, $\tilde\zeta$ finite,
\[
 D(\zeta,\tilde\zeta) = \prod_{i,\alpha:\tilde\zeta(i,\alpha)=1} \zeta(i,\alpha)
\]
then we have the duality relation from \cite{Liggett}
\be\label{ladderdual}
\E_\zeta D(\zeta_t,\tilde\zeta) = \E_{\tilde\zeta} D(\zeta,\tilde\zeta_t)
\ee
where $\zeta_t$, $\tilde\zeta_t$ are independent copies of the ladder SEP with
generator \eqref{laddersep} starting from $\zeta$, resp.\ $\tilde\zeta$.
\ep
\subsection{From the ladder SEP to the $2j$-SEP}
To define the $2j$-SEP on a graph ${\cal S}$ we consider,
for a given SEP on a ladder graph ${\cal S}\times {\cal I}$ with
$2j$ levels, the process which consists of giving at time $t>0$, and each site $i\in {\cal S}$
the number of levels $(i,\alpha)$ which are occupied in $\zeta_t$.
More precisely, define the map $\pi: \Omega\to \Omega^{(j)} = \{0,1,\ldots, 2j\}^{\cal S}$
\be\label{redulad}
\zeta\mapsto \pi (\zeta)  = \eta \qquad \text{with} \quad \eta_i = \sum_{\alpha\in {\cal I}} \zeta(i,\alpha)
\ee
Then we have the following theorem.
\bt\label{sepjthm}
Let $\zeta_t$ be the ladder sep with generator \eqref{laddersep}. Then the following holds:
\begin{itemize}
\item[a)]

$\eta_t = \pi(\zeta_t)$ is a Markov process on $\Omega^{(j)}$ with generator
\be\label{sepj}
L^{(j)} f(\eta) = \sum_{i,l\in S} p(i,l) \eta_i (2j-\eta_l) (f(\eta^{i,l} ) -f(\eta))
\ee
This process will be called the $2j$-SEP or reduced ladder SEP with $2j$ levels.
\item[b)] The process $\eta_t$ with generator $L^{(j)}$ is self-dual with duality function
\be\label{dualj}
D(\eta,\xi) = \prod_{i\in S} \frac{{\eta_i\choose\xi_i}}{{2j\choose\xi_i}}
\ee
for $\xi\leq \eta$ a configuration with a finite number of
particles ($D$ is defined to be zero in other cases). More precisely, we have
\be\label{dualityrel}
\E_\eta D(\eta_t,\xi ) = \E_\xi D(\eta, \xi_t)
\ee
\item[c)] The extremal invariant measures are of the form
\[
 \nu^{(j)}_\rho = \otimes_{i\in S} {\rm Bin} (2j, \rho_i)
\]
where $\rho_i$ is harmonic for $p(i,l)$, i.e., such that
\[
 \sum_l p(i,l)\rho_l= \rho_i
\]
In particular, if the only bounded harmonic functions are constants, then
the only extremal invariant measures are products of binomials with constant
density.
\end{itemize}
\et
\bpr
For point (a) remark that the jump rates in the generator
\eqref{sepj} only depend on the number of particles at
a site, and not on the levels. Therefore, if
$f:\Omega\to\R$ depends on $\zeta$ only through $\eta=\pi(\zeta)$, i.e.,
if $f(\zeta) = \psi(\pi(\zeta))=  \psi(\eta)$, then
\be\label{redgen}
L f (\zeta) = L^{(j)} \psi (\eta)
\ee
Therefore, for every local function $\psi:\Omega^{(j)}\to\R$,
\be\label{mart}
\psi (\pi(\zeta_t) ) -\psi(\pi(\zeta_0))- \int_{0}^t L^{(j)} (\psi) (\pi(\zeta_s))ds =M_t
\ee
is a martingale w.r.t.\ the filtration
$\fe_t= \si\{\pi(\zeta)_s: 0\leq s\leq t\}$.
This shows that $\pi(\zeta_t)$ is a solution of the martingale problem
associated to the generator $L^{(j)}$, and hence coincides
with the unique Markov process generated by $L^{(j)}$ (see Th. 4.1, page 182, of \cite{trotter-kurtz}).

Now we turn to point (b). At each site $i\in S$ we choose
$\xi_i$ levels at random. For a given
configuration $\eta\in \Omega_j$, we choose
$\zeta\in\Omega$ consistent, i.e., such that $\pi (\zeta) = \eta$.
Then the probability (w.r.t. to the random choices)
that all chosen levels are occupied in $\zeta$ is exactly
equal to $D(\eta,\xi)$. As $\pi (\zeta_t) =\eta_t$, the probability
that the chosen levels are occupied at time $t$ (i.e., in $\zeta_t$) is given by
$\E_\eta D(\eta_t,\xi)$. By self-duality of the ladder SEP
(Prop. \ref{old}), the event
that at time $t>0$ the chosen levels are occupied is
the same as the probability that the particles evolving
from the chosen levels during a time $t$ find themselves
at positions which are occupied in $\zeta$. The latter
probability equals $\E_\xi D(\xi_t,\eta)$.

Point c) follows from the fact that for the ladder SEP with generator
\eqref{laddersep}, the product Bernoulli measures indexed by harmonic functions
of $p(i,j)$ are the extremal invariant measures ( see \cite{Liggett}
for details) and the image measure of a product of Bernoulli measure
under $\pi$ is a product of Binomial measures.
\epr
\subsection{Limiting processes as $j\to\infty$}
In this section we show that for large $j$ the
$2j$-SEP converges, when considered on an appropriate time scale,
either to a system of independent random walkers or to a deterministic limit,
depending on the initial density.
We remind the reader that for independent random walkers on a graph {\cal S},
the configuration space is $\Omega_\infty = \N^{\cal S}$ and the generator
is given by
\be\label{indgen}
L_{irw} f(\eta) = \sum_{i,l\in {\cal S}}\eta (i) p(i,l) (f(\eta^{i,l})-f(\eta))
\ee
The stationary measures are products of Poisson measures, and the process
with generator \eqref{indgen} is self-dual with self-duality function
\be\label{poispol}
D_{irw}(\eta,\xi) = \prod_{i\in S} \frac{\eta_i !}{(\eta_i-\xi_i)!}
\ee
for finite configurations $\xi\leq \eta$, and $D=0$ otherwise.

The relation with the reduced ladder SEP for large $j$ is given in
the following theorem. \bt \label{jinf} Consider the process
$\{\eta^{(j)}_{t}:t\geq 0\}$ with generator \eqref{sepj} started
from initial configuration $\eta^{(j)} \in \Omega^{(j)}$. Suppose
that, as $j\to\infty$, $\eta^{(j)}\to \eta\in \Omega_\infty$, then
the process $\{\eta^{(j)}_{t/2j}:t\geq 0\}$ converges weakly in path space
to a system of
independent random walkers with generator \eqref{indgen} and initial
configuration $\eta$. \et \bpr The process
$\{\eta^{(j)}_{t/2j}:t\geq 0\}$ has generator \be\label{jgen}
\loc'_j = \frac{1}{2j} L^{(j)} \ee In order to have a sequence of
processes all defined on the same sample space $\Omega^{(\infty)}$
we consider the auxiliary process on $\N^S$ with generator \be
\label{auxgen} \loc''_j f(\eta) = \sum_{i,l\in S}p(i,l) \eta_i
\left(1-\frac{\eta_l}{2j}\right)I (\eta_i\leq 2j) I (\eta_l\leq 2j)
\left(f(\eta^{i,l}) - f(\eta)\right) \ee This auxiliary process
behaves as the process with generator $\loc'_j$ except for sites
which have more than $2j$ particles, which are frozen. Started from
an initial configuration with all sites having at most $2j$
particles, this process coincides with the process $\eta_{t/2j}$.
For any local function $f:\N^S\to\R$, we then have
\begin{eqnarray}
\label{trot}
\lim_{j\to\infty}\loc''_j f(\eta)
&=&
\lim_{j\to\infty} \sum_{i,l\in S} p(i,l)\eta_i
\left(1-\frac{\eta_l}{2j}\right)I (\eta_i\leq 2j) I (\eta_l\leq 2j)
\left(f(\eta^{i,l}) - f(\eta)\right) \nonumber \\
&=&
\loc_{irw} f(\eta)
\end{eqnarray}
Therefore, by the Trotter-Kurtz theorem (see Theorem 2.5 of \cite{trotter-kurtz}),
this implies that the corresponding processes
$\eta_{t/2j}$ converge weakly on path space as $j\to\infty$ to the process
with generator $\loc_{irw}$.
\epr
To see that \eqref{poispol} is (up to a multiplicative consant)
a limit of duality functions
of the $2j$-SEP, we start from the symmetry \eqref{cc1}
and use the similarity transformation with $Q^{(j)}$
given by
$$
Q^{(j)}(\eta,\xi) = \prod_{i\in {\cal S}} Q^{(j)}_i(\eta_i,\xi_i)
$$
with
$$
Q^{(j)}_i(\eta_i,\xi_i) = \delta_{\eta_i,\xi_i} {2j \choose \eta_i}
\left(\frac{1}{2j}\right)^{\eta_i} \left(1- \frac{1}{2j}\right)^{2j -\eta_i}
$$
Then the duality functions
$$
\tilde D^{(j)} = {Q^{(j)}}^{-1} S
$$
with $S$ defined in \eqref{ppp},
satisfy
$$
\lim_{j\to\infty} \tilde D^{(j)} =  e D_{irw}\;.
$$

Another possible limit is obtained when the initial condition
has a number of particles that diverges proportional to $j$.
This limit, as can be understood from the law of large numbers,
is deterministic.
\bt
Consider the process
\[
 x^{(j)}_i (t) = \frac{\eta^{(j)}_{t/2j}}{2j}
\]
and suppose that $x^{(j)}_i(0)\to x_i(0)\in [0,1]$ for all
$i\in {\cal S}$ as $j\to\infty$.
Then we have
that $x^{(j)}_i(t)$ converges to a deterministic system
of coupled differential equations with generator
\be\label{detgen}
L f(x) = \sum_{i,l\in S} p(i,l) x_i (1-x_l) \left(\frac{\partial}{\partial x_l}- \frac{\partial}{\partial x_i}\right)
\ee
\et
\bpr
The generator of the process $x_i^{(j)} (t)$ reads
\[
 \loc'_j f(x) = 2j\sum_{i,l} p(i,l) x_i (1-x_l) (f(x^{(j); i,l} ) -f(x^{(j)}))
\]
where $x^{(j),i,l}$ arises from $x^{(j)}$ by removing a unit $1/2j$ from $ i\in {\cal S}$
and putting it at $l\in {\cal S}$.
Therefore, for a local smooth function $f: [0,1]^S \to\R$ we have, by Taylor expansion
\[
 \loc'_j f(x)= \sum_{i,l} p(i,l) x_i (1-x_l) \left(\frac{\partial}{\partial x_l}- \frac{\partial}{\partial x_i}\right)
f(x) + O\left(\frac{1}{2j}\right)
\]
The result then follows once more from
the Trotter Kurtz theorem.
Since the limiting generator is a first order differential operator,
the corresponding process is deterministic.
\epr
\subsection{Boundary driven case}
We first discuss a duality theorem for standard (i.e. $j=1/2$)
symmetric exclusion with extra creation and annihilation of particles at the boundary.
The context is a countable set ${\cal S}$, of which we distinguish
a subset $\partial {\cal S}\subset {\cal S}$ called the boundary.
We then consider the generator
\begin{eqnarray}\label{boundgen}
\loc f(\eta) &=& \sum_{i,l\in {\cal S}}p(i,l) (\eta(i)(1-\eta(l))
(f(\eta^{i,l})- f(\eta))
\nonumber\\
&+& \sum_{i\in\partial {\cal S}} (1-\rho_i) \eta (i)
\left(f(\eta^i)-f(\eta)\right) +  \rho_i (1-\eta(i))
\left(f(\eta^i)-f(\eta)\right)
\end{eqnarray}
where $0<\rho_i <1$ represent the densities of the
particle reservoirs with which the system
is in contact at the boundary sites, and where $\eta^i$ is the configuration
obtained from $\eta$ by flipping the occupancy at $i$.

The first part of the generator represents the hopping
of particles on ${\cal S}$ according to a symmetric exclusion process,
whereas the second part represents creation and annihilation
of particles at the boundary sites.

To introduce duality for this process, we introduce a set
$\partial_e {\cal S}$ of sink sites, and a bijection
$i\mapsto i_e$ which associates each site $i\in\partial {\cal S}$
to a sink site.
The dual process will then be a process that
behaves as the exclusion process in the bulk, but
particles can leave the system via boundary sites to sink sites, and
will then be stuck at sink sites.
More precisely, a configuration of the dual process
is a map
\[
 \xi: {\cal S}\cup \partial_e {\cal S} \to \N
\]
such that $\xi(i)\in \{0,1\}$ for $i\in {\cal S}$ (only the sink sites can
contain more than one particle). We call $\Omega_{dual}$ the set of
all configurations of the dual process.

The duality function is defined
as follows: for $\eta\in \{ 0,1\}^{{\cal S}}$, $\xi\in \Omega_{dual}$
\be\label{bounddual}
D(\eta,\xi) = \left(\prod_{i\in {\cal S}:\xi(i)=1} \eta(i)\right)\left(\prod_{i\in\partial_e {\cal S}} \rho_i^{\xi_i}\right)
\ee

The idea here is that we have the ordinary duality function for
the sites $i\in {\cal S}$ and for the sink sites, we replace
the variable $\eta_i$ by its expectation $\rho_i$, corresponding
to the stationary measure of the boundary generator $L_i$.

The generator of the dual process is then defined as follows:
\be\label{dualboundgen} \loc_{dual} f(\xi) = \sum_{i,l\in {\cal
S}}p(i,l) (\xi(i)(1-\xi(l)) (f(\xi^{i,l})- f(\xi)) + \sum_{i\in
\partial {\cal S}} \xi(i) \left(f(\xi^{i,i_e})- f(\xi)\right) \ee

We then have \bt\label{dualboundthm} The boundary driven exclusion
process $(\eta_t)_{t\ge 0}$ with generator ${\cal L}$ in
\eqref{boundgen} and the process $(\xi_t)_{t\ge 0}$ with generator
${\cal L}_{dual}$ in \eqref{dualboundgen} are dual with duality
function $D(\eta,\xi)$ given by \eqref{bounddual}, i.e.,
\be\label{boundexdual} \E_\eta D(\eta_t,\xi) = \E_{\xi}
D(\eta,\xi_t) \ee \et \bpr Abbreviate \be L_i f (\eta) =(1-\rho_i)
\eta (i) \left(f(\eta^i)-f(\eta)\right) +  \rho_i (1-\eta(i))
\left(f(\eta^i)-f(\eta)\right) \ee and \be L^{dual}_i f(\xi) =
\xi(i) \left(f(\xi^{i,i_e})- f(\xi)\right) \ee For $f(\eta)=\eta(i)$
one sees that \be L_i f(\eta)= \rho_i-\eta (i) \ee and hence for
$\xi$ a dual configuration which is non-zero only on the sites
$i\in\partial {\cal S}$ and on the corresponding sink site
$i_e\in\partial_e {\cal S}$, we find
\begin{eqnarray*}
L_i D(\eta, \xi)&=& \rho_i^{\xi_{i_e}} \left(L_i\left(\eta(i) \xi(i)
+ (1-\xi(i))\right)\right)
\\
&=& \xi(i) \rho_i^{\xi_{i_e}} (\rho_i-\eta(i))
\\
&=& \xi(i)\left(\rho_i^{\xi_{i_e}
+1}-\rho_i^{\xi_{i_e}}\eta(i)\right) = L^{dual}_i D(\xi,\eta)
\end{eqnarray*}
From that and the self-duality of the symmetric exclusion process,
it follows that
\be
\loc D(\eta, \xi) = \loc_{dual} D(\eta, \xi)
\ee
\epr
In order to apply this duality result for the boundary driven process with
generator \eqref{sepj}, we first look at the boundary driven exclusion
process on a ladder graph.
More precisely, for $\zeta\in \{0,1\}^{{\cal S}\times{\cal I}}$ we consider the process
$(\zeta_t)_{t\ge0}$ with generator
\begin{eqnarray}
\label{boundgenladder}
\loc f(\zeta) &=& \sum_{i,l\in {\cal S}}\sum_{\alpha,\beta\in {\cal I}}p(i,l) (\zeta(i,\alpha)(1-\zeta(l,\beta))
(f(\zeta^{(i,\alpha),(l,\beta)})- f(\zeta))
\\
&+& \sum_{i\in\partial {\cal S}}\sum_{\alpha\in {\cal I}} (1-\rho_i) \zeta (i,\alpha) \left(f(\zeta^{(i,\alpha)})-f(\zeta)\right)
+  \rho_i (1-\zeta(i,\alpha)) \left(f(\zeta^{(i,\alpha)})-f(\zeta)\right)\nonumber
\end{eqnarray}
In words, this process is the ladder SEP, with additional boundary driving,
where the creation and annihilation rate of particles at the boundary
sites does not depend on the level.
If we consider the reduced process, consisting
of counting at each site $i\in {\cal S}$ how many levels
in ${\cal I}$ are occupied, i.e. $\eta_t = \pi(\zeta_t)$
then we recover once again a Markov process (cfr. Theorem \ref{sepjthm}).
This process, defined on the state space $\Omega^{(j)} = \{0,1,\ldots 2j\}^{{\cal S}}$
and called the boundary driven $2j$-SEP, has generator
\begin{eqnarray}
\label{come}
\loc_j f(\eta) &=& \sum_{i,l\in {\cal S}}p(i,l) (\eta(i)(2j-\eta(l))
(f(\eta^{i,l})- f(\eta))
\nonumber\\
&+& \sum_{i\in\partial {\cal S}} (1-\rho_i) \eta (i) \left(f(\eta^{(i,\alpha)})-f(\eta)\right)
+  \rho_i (2j-\eta_i) \left(f(\eta^i)-f(\eta)\right)
\end{eqnarray}
It turns out that the boundary driven $2j$-SEP has a nice dual too.
To introduce this dual, we consider admissible dual configurations as
maps
$\xi:{\cal S}\cup\partial_e {\cal S}\to \N$ such that $0\leq \xi (i)\leq 2j$ for $i\in {\cal S}$
(only the sink sites can contain more that $2j$ particles).
The generator of the dual of the boundary driven $2j$-SEP is a process
on admissible dual configurations, defined by
\begin{eqnarray}\label{dualJboundarygen}
\loc^{dual}_j f(\xi) &=& \sum_{i,l\in {\cal S}}p(i,l) (\xi(i)(2j-\xi(l))
(f(\xi^{i,l})- f(\xi))
\nonumber\\
&+& \sum_{i\in\partial {\cal S}}  \xi (i) \left(f(\xi^{i,i_e})-f(\xi)\right)
\end{eqnarray}
From Theorem \ref{dualboundthm} we then infer, in the same way as we
derived Theorem \ref{sepjthm} the following.
\bt\label{laddersepboundthm} Let $(\eta_t)_{t\ge 0}$ denote the
boundary driven $2j$-SEP with generator \eqref{come}. Then
$(\eta_t)_{t\ge 0}$ is dual to the process $(\xi_t)_{t\ge 0}$ with
generator \eqref{dualJboundarygen} with duality function given by
\be\label{boundaryladderdual} D(\eta,\xi) = \prod_{i\in\partial_e
{\cal S}} \rho_i^{\xi_{i}}\prod_{i\in {\cal S}}
\frac{{\eta_i\choose\xi_i}}{{2j\choose\xi_i}} \ee \et
\bpr Denote
for $i\in\partial {\cal S}$,
\[
\loc_{i} f(\eta)= (1-\rho_i) \eta (i)
\left(f(\eta^{(i,\alpha)})-f(\eta)\right) +  \rho_i (2j-\eta(i))
\left(f(\eta^i)-f(\eta)\right)
\]
and
\[
 \loc_{i}^{dual} f(\xi) = \xi (i) \left(f(\xi^{i,i_e})-f(\xi)\right)
\]
One then easily computes that
for $\xi$ a dual configuration
which is non-zero only on the sites $i\in\partial {\cal S}$ and
on the corresponding sink site $i_e\in\partial_e {\cal S}$,
\begin{eqnarray*}
 \loc_{i} D(\eta, \xi)
&=&\rho_i^{\xi_{i_e}}\frac{1}{{2j \choose \xi_i}} \loc_{i} \left(
{\eta_i \choose \xi_i}\right)
\\
&=& \rho_i^{\xi_{i_e}}\frac{1}{{2j \choose \xi_i}} \left[
(1-\rho_i)\eta_i \left({\eta_i-1 \choose \xi_i}-{\eta_i \choose
\xi_i}\right) + (2j-\eta_i) \rho_i \left({\eta_i+1 \choose
\xi_i}-{\eta_i \choose \xi_i}\right) \right]
\\
&=& \xi_i (\rho_i^{\xi_{i_e}+1}-\rho_i) \frac{{\eta_i \choose
\xi_i}}{{2j \choose \xi_i}} =\loc_{i}^{dual} D(\eta,\xi)
\end{eqnarray*}
The result then follows from combination of
this fact and the duality relation \eqref{dualityrel}.
\epr

\section{The Brownian Momentum Process and $SU(1,1)$ symmetry}
\label{secmep}

In this section we study the Brownian momentum process
that was introduced in \cite{gk,olla}. We will recover
duality \cite{gkr} in the context of our main Theorems
and study the reversible measures of the dual process.

\subsection{Generator and Quantum spin chain}
Let ${\cal S}$ be a countable set and $p(i,j)$ a symmetric random walk
transition probability on ${\cal S}$. The Brownian momentum process
is a Markov process $(x_t)_{t\ge 0}$ on $\R^{\cal S}$, with generator
\be\label{bepgeno}
L = \sum_{i,j\in {\cal S}} p(i,j) L_{ij}
\ee
where
\be\label{bepgenij}
L_{ij} = \lij
\ee
and $p(i,j)$ is a symmetric random walk kernel on ${\cal S}$.

The generator $L_{ij}$ conserves the energy $x_i^2 + x_j^2$
and generates a Brownian rotation of the angle $\theta_{ij} = \arctan(x_j/x_i)$.
The interpretation of the generator \eqref{bepgeno} is then as follows:
each bond independently, at rate $p(i,j)$ undergoes a Brownian
rotation of its angle
$\theta_{ij} = \arctan(x_j/x_i)$. An important example to keep in mind
is ${\cal S}=\Zd$, and $p(i,j)$ the nearest neighbor symmetric random walk.

The processes with generator $L$ can be related to quantum spin chains \cite{gkr}.
Consider the  operators
\begin{eqnarray}\label{kopa11}
K^+_{i} &=& \frac12 x_{i}^2\nonumber\\
K_{i}^- &=& \frac12 \frac{\partial^2}{\partial x_{i}^2}\nonumber\\
K_{i}^0 &=& \frac14 \left(\frac{\partial}{\partial x_{i}}(x_{i} \cdot) +x_{i}\frac{\partial}{\partial x_{i}}\right)
\end{eqnarray}
which satisfy the commutation relations of $SU(1,1)$:
\begin{eqnarray}\label{su11com2}
[K_i^0, K_i^\pm] &=& \pm K_i^\pm \nonumber\\
{[} K_i^-, K_i^+] &= & 2 K_i^0
\end{eqnarray}
Then the negative of the adjoint of the generator $L$
can be seen as the quantum ``Hamiltonian''
\be\label{bepham11}
H = -L^* = -4\sum_{ij\in {\cal S}} p(i,j)
\left( K_i^{+} K_j^{-} + K_i^{-} K_j^{+}
-
2K^{0}_i K_j^{0} + \frac{1}{8} \right)
\ee
with spin satisfying the $SU(1,1)$ algebra
(in a representation with spin value $1/4$).

\subsection{Dual process}
In \cite{gkr} we showed that the process with generator $L$ in
\eqref{bepgeno} and \eqref{bepgenij} has a dual, which is a system
of interacting random walkers on ${\cal S}$. We show here how this
dual process comes out of the structure of the Hamiltonian
\eqref{bepham11}.

We notice that the $SU(1,1)$ group admits a discrete (infinite dimensional) representation
as (unbounded) operators on $l_2 (\N)$:
\begin{eqnarray}\label{kopdi}
\caK_i^+ |\xi_i\rangle &=& \left(\frac12 + \xi_i\right) |\xi_i+1\rangle
\nonumber\\
\caK_i^- |\xi_i\rangle &=& \xi_i |\xi_i-1\rangle
\nonumber\\
\caK_i^0 |\xi_i\rangle &=& \left( \xi_i + \frac14\right) |\xi_i\rangle
\end{eqnarray}
where $i\in {\cal S}$ and  $\xi_i\in\N$ and
$|0\rangle,|1\rangle,|2\rangle,\ldots$ denotes the canonical basis
on $l_2 (\N)$. It is immediately checked that the (unbounded)
operators in \eqref{kopdi} satisfy the $SU(1,1)$ commutation
relations in \eqref{su11com2}. We then define a new generator via
the same Hamiltonian as in \eqref{bepham11}, but now in the
representation \eqref{kopdi}: \be\label{bepdualgen} H_{dual} =
-L_{dual}^* = -4\sum_{ij\in {\cal S}} p(i,j) \left(\caK_i^+ \caK_j^-
+\caK_i^- \caK_j^+ - 2\caK^0_i\caK_j^0 + \frac18\right) \ee From the
previous equation and using the representation \eqref{kopdi} we
deduce that the Hamiltonian above defines a Markov process
$(\xi_t)_{t\ge 0}$ with state space $\N^{{\cal S}}$ and generator
\be\label{dualbepgen} L_{dual} =\sum_{i,j\in {\cal S}} p(i,j)
L^{dual}_{ij} \ee where \begin{eqnarray}\label{eldualbepgen}
L^{dual}_{ij} f(\xi)  & = & 2\xi_i (2\xi_j+1) (f(\xi^{i,j})-f(\xi))
+ \nonumber\\ & & 2\xi_j (2\xi_i+1) (f(\xi^{j,i})-f(\xi))
\end{eqnarray} with $\xi^{i,j}$ the configuration obtained from $\xi$
by removing a particle from $i$ and putting it at $j$. Note that, in
general, changing a representation does not imply that a generator
continues to be a generator: the fact that $H$ and $H_{dual}$ are
well-defined as a Hamiltonian is conserved by similarity transformations
(change of representation) but their property of being (minus)
the adjoint of the generator of a Markov process is dependent on the
representation, and needs to be verified by hand.

\subsection{The duality function explained}\label{dualex}
In \cite{gkr} we found that $L$ and $L_{dual}$ are dual processes,
with duality function \be D(x,\xi) = \prod_{i\in {\cal S}} D_i(x_i,
\xi_i) \ee with \be D_i(x_i, \xi_i) =
\frac{x_i^{2\xi_i}}{(2\xi_i-1)!!} \ee We now show how these
functions arise from the change of representation. Suppose that we
find a function
$$
C=C(x,\xi) = \prod_{i\in {\cal S}} C_i(x_i, \xi_i)
$$
such that \be\label{borges} K_i^{a} C_i =  C_i \caK_i^{a} \ee for
$a\in \{ +,-,0\}$, $i\in{\cal S}$ and $K_i^{a}$, resp.\ $\caK_i^a$
defined in \eqref{kopa11}, \eqref{kopdi}. The ``matrix product'' in
the lhs of \eqref{borges} is defined as the differential operator
$K_i^a$ working on the $x_i$-variable of $C_i(x_i,\xi_i)$, and in
the rhs $C_i \caK_i^a(x_i,\xi_i) = \sum_{\xi_i'} C(x_i,\xi_i')
\caK_i^a (\xi_i',\xi_i)$.

Then for the generators $L$ in \eqref{bepgeno},\eqref{bepgenij} and
the generator $L_{dual}$ in \eqref{dualbepgen}, \eqref{eldualbepgen}
we find, as a consequence of \eqref{borges} and using that $L^*= L$,
\be L C = L^* C = CL_{dual}^* \ee i.e., such a function $C$ is a
duality function (cfr. \eqref{dualmagen}).

The equation \eqref{borges} is most easy of $a= +$, it then reads
\be\label{murakami} \frac12 x_i^2 C_i(x_i,\xi_i) = \left(\frac12
+\xi_i \right) C_i(x_i, \xi_i+1) \ee To find $C_i(x_i,0)$ we use
$K_i^- C_i(x_i,0) =0$ (that follows from
\eqref{kopdi}, \eqref{borges}) so we can choose $C_i(x_i,0)=1$ and then find,
via \eqref{murakami} \be\label{kafka}
 C_i(x_i,\xi_i) = \frac{x_i^{2\xi_i}}{(2\xi_i-1)!!}
\ee which is exactly the duality function that we found in
\cite{gkr}. It is then easy to verify that \eqref{borges} is also
satisfied for $a\in\{-, 0\}$ with the choice \eqref{kafka}.

\subsection{Reversible measures  of the dual-BEP}

The dual of the BEP,  with generator $L_{dual}$ in
\eqref{dualbepgen} and \eqref{eldualbepgen}, is in itself an
interacting particle system (particles attract each other), and it can
therefore be considered as a model of independent interest. In some
sense, it can be viewed as ``the bosonic counterpart'' of the SEP.
Surprisingly, despite the interaction, the process has reversible
product measures, as is shown below. Remark that, due to the
attractive interaction between the particles, this process does not
fall in the class of ``misantrope processes'' considered in
\cite{ravi}, \cite{coc} (where one also has in particular cases
stationary product measures, despite interaction). \bt\label{bosmes}
Consider, for $\lambda < 1/2$ the translation invariant product
measure $\nu_\lambda$ on $\N^{\cal S}$ with marginals
\be\label{revprod} \nu_\lambda (\eta_0= k) =
\frac{1}{Z_\lambda}\frac{(2k-1)!!}{k!} \lambda^k \ee where the
normalization is \be\label{zet} Z_\lambda =
\sum_{k=0}^\infty\frac{(2k-1)!!}{k!} \lambda^k =
\frac{1}{\sqrt{(1-2\lambda )}}\;. \ee Then $\nu_\lambda$ is
reversible for the process with generator $L_{dual}$ in
\eqref{dualbepgen} and \eqref{eldualbepgen}. \et \bpr From the
generator \eqref{dualbepgen}, \eqref{eldualbepgen}, we infer that
\[
\alpha_i\alpha_j 2i (2j+1) = 2(j+1) (2i-1)\alpha_{j+1}\alpha_{i-1}
\]
is a sufficient condition for detailed balance of a product measure
$\mu$ with marginals $\mu (\eta_0=k) =\alpha_k$ for the generator $
L^{dual}_{ij}$, which is sufficient for detailed balance for
$L_{dual}$. This leads to
\[
 \frac{\alpha_{j+1}}{\alpha_j} = \frac{2j+1}{2j+2} (2c)
\]
for some positive constant $c$.
This in turn gives
\[
 \alpha_j = \frac{(2j-1)!!}{j!} (2c)^j \alpha_0
\]
which is \eqref{revprod}.
The explicit expression for $Z_\lambda$ can be obtained easily from the identity
\[
 \int_{-\infty}^\infty x^{2k} e^{-x^2/2} dx= \sqrt{2\pi} (2k-1)!!
\]
\epr

\section{The Brownian Energy Process}

As it was done for SEP, it is interesting to consider the Brownian
Momentum Process on ladder graph ${\cal S}\times{\cal I}$ with
$|{\cal I}|=m \in\N $ levels and look at the induced process which
gives the energy at each site.

\subsection{Generator}
Consider the generator, working on local functions $f:\R^{{\cal
S}\times {\cal I}}\to\R$ \be\label{bepgenm} L = \sum_{i,j\in{\cal
S}}\sum_{\alpha, \beta=1}^m p(i,j) L_{i\alpha ,j\beta} \ee where
\be\label{bepgenija} L_{i\alpha ,j\beta}= \lija \;.\ee In this
section we show that for the process with the generator above, the
total energy per site defined via \be\label{enm}
 z_i = \sum_{\alpha=1}^m x_{i,\alpha}^2
\ee is again a Markov process \bt Consider the process $x(t) =
(x_{i,\alpha} (t))_{i\in {\cal S}, \alpha= 1,\ldots, m}$ with
generator $L$ of \eqref{bepgenm} and \eqref{bepgenija}. Consider the
corresponding process $z(t) =( z_i (t))_{i\in {\cal S}}$ defined via
the mapping \eqref{enm}. This is a Markov process on $\R_{+}^{{\cal
S}}$ with generator \be\label{bepengen} L^{(m)} = \sum_{ij\in {\cal
S}} p(i,j) L^{(m)}_{ij} \ee with \be\label{bepengenij}
L^{(m)}_{ij}=4 z_i z_j\left(\frac{\partial}{\partial z_i}
-\frac{\partial}{\partial z_{j}}\right)^2 - 2m (z_i-z_{j})
\left(\frac{\partial}{\partial z_i} - \frac{\partial}{
\partial z_{j}}\right) \ee and with stationary measures
which are product measures with chi-squared marginals.\et \bpr
Denote $\pi: (x_{i,\alpha})_{i\in {\cal S}, \alpha= 1,\ldots,
m}\mapsto (z_i)_{i\in {\cal S}}$. Denote by $\partial_i$ partial
derivative w.r.t.\ $z_i$ and by $\partial_{i,\alpha}$ partial
derivative w.r.t.\ $x_{i,\alpha}$. Then, using the identities
\begin{eqnarray}
\label{ident}
 \partial_{i,\alpha} &=& 2 x_{i,\alpha}\partial_i
\nonumber\\
\partial_{i,\alpha}^2 &=& 2\partial_i + 4 x_{i\,\alpha}^2 \partial^2_{i}
\nonumber\\
\partial_{i,\alpha}\partial_{j,\beta} &=& 4x_{i,\alpha}x_{j,\beta} \partial_{i}\partial_{j}
\end{eqnarray}
we find for a function $f:\R^{{\cal S}\times {\cal I}}\to \R$
depending on $x$ only through $z=\pi(x)$ \be\label{dosto} L
f\circ\pi (x) = (L^{(m)} f) (\pi (x)) \ee The proof then proceeds
via the martingale problem as in the proof of Theorem \ref{sepjthm}.
The stationary measure of the process with generator $L^{(m)}$ is
deduced from the knowledge of the stationary measure for the process
with generator $L$. Indeed, it is easy to check that for the process
$x_{i,\alpha} (t)$, products of Gaussian measures $\otimes_{i\in
{\cal S}_1,\alpha} N(0,\si^2)$ are invariant and ergodic. The image
measure under the transformation $\pi (x) =z$ are products
$\otimes_{i\in {\cal S}_1} \chi^2_m (\si)$ where for $\si=1$,
$\chi^2_m (1)$ is the chi-squared distribution with $m$ degrees of
freedom, and for general $\si$, follows from scaling $\chi^2_m
(\si^2) = \si^2 \chi^2_m (1)$. \epr

\subsection{Duality}
In this section we show that the BEP defined above has a dual
process which is again a jump process.

To construct the dual we follow a procedure similar to the one of
the previous section. Remark that the generator $L$ in
\eqref{bepgenm} and \eqref{bepgenija} can be written in terms of the
operators \be\label{kam} K_i^{a, (m)} = \sum_{\alpha=1}^m
K^{a}_{i,\alpha} \ee with $a\in \{ +,-, 0\}$, where
\begin{eqnarray}\label{kopa}
K^+_{i,\alpha} &=& \frac12 x_{i,\alpha}^2\nonumber\\
K_{i,\alpha}^- &=& \frac12 \frac{\partial^2}{\partial x_{i,\alpha}^2}\nonumber\\
K_{i,\alpha}^0 &=& \frac14 \left(\frac{\partial}{\partial x_{i,\alpha}}(x_{i,\alpha} \cdot) +x_{i,\alpha}\frac{\partial}{\partial x_{i,\alpha}}\right)
\end{eqnarray}
In other words $L$ is related to a quantum spin chain with
Hamiltonian \be\label{bepham} H^{(m)} = -L^*= -4\sum_{ij\in {\cal
S}} p(i,j) \left( K_i^{+, (m)} K_j^{-,(m)} + K_i^{-,(m)} K_j^{+,
(m)} - 2K^{0, (m)}_i K_j^{0, (m)} + \frac{m^2}{8} \right) \ee where
the $K$-operators in \eqref{kam} and \eqref{kopa} satisfy the
commutation relations of $SU(1,1)$. Moreover the $K$ operators
defined in \eqref{kam}, \eqref{kopa}, can be rewritten in
$z$-variables as follows:
\begin{eqnarray}\label{kaz}
K_i^{+, (m)} &=& \frac12 z_i \nonumber\\
K_i^{- (m)} &=& 2z_i \partial_i^2 + m \partial_i\nonumber\\
K_i^{0,(m)} &=&  z_i \partial_i + \frac{m}{4}
\end{eqnarray}
The generator $L^{(m)}$ of \eqref{bepengen} is then simply minus the
adjoint of the Hamiltonian $H^{(m)}$ in \eqref{bepham}, rewritten
with $K$-operators in $z$-variables.

At this point it is important to remark that the $SU(1,1)$ group
admits a family of discrete infinite dimensional discrete
representation labeled by $m\in\N$ given by
\begin{eqnarray}\label{kopdim}
\caK_i^{+,(m)} |\xi_i\rangle &=& \left(\frac{m}{2} + \xi_i\right)
|\xi_i+1\rangle
\nonumber\\
\caK_i^{-,(m)} |\xi_i\rangle &=& \xi_i |\xi_i-1\rangle
\nonumber\\
\caK_i^{0,(m)} |\xi_i\rangle &=& \left(\frac{m}{4} + \xi_i\right)
|\xi_i\rangle
\end{eqnarray}
where $i\in {\cal S}$ and  $\xi_i\in\N$ and
$|0\rangle,|1\rangle,|2\rangle,\ldots$ denotes the canonical basis
on $l_2 (\N)$. We then define a new generator via the same
Hamiltonian as in \eqref{bepham}, but now in the representation
\eqref{kopdim}: \be\label{bepdualgenm} H_{dual}^{(m)} = -L_{dual}^*
= -4\sum_{ij\in {\cal S}} p(i,j) \left(\caK_i^{+,(m)} \caK_j^{-,(m)}
+\caK_i^{-,(m)} \caK_j^{+,(m)} - 2\caK^{0,(m)}_i\caK_j^{0,(m)} +
\frac{m^2}{8}\right) \ee Using the representation \eqref{kopdim} we
deduce that the Hamiltonian above defines a Markov process
$(\xi_t)_{t\ge 0}$ with state space $\N^{{\cal S}}$ and generator
\be\label{dualbepgenm} L^{(m)}_{dual} =\sum_{i,j\in {\cal S}} p(i,j)
L^{(m), dual}_{ij} \ee where \begin{eqnarray}\label{eldualbepgenm}
L^{(m), dual}_{ij} f(\xi)  & = &   2\xi_i (2\xi_j+m)
(f(\xi^{i,j})-f(\xi)) + \nonumber\\ &  &2\xi_j (2\xi_i+m)
(f(\xi^{j,i})-f(\xi))\end{eqnarray}

The duality function are given in the following theorem.
\bt\label{ult} The processes with generator $L^{(m)}$ and
$L^{(m)}_{dual}$ are each others dual, with duality function
\be\label{tolstoj} D(z,\xi) = \prod_{i\in{\cal S}} D_i(z_i,\xi_i)
\ee where \be\label{gogol} D_i(z_i,\xi_i)= z_i^{\xi_i} \frac{\Gamma
\left(\frac{m}{2}\right)}{2^{\xi_i} \Gamma
\left(\frac{m}{2}+\xi_i\right)} \ee with $\Gamma (t) = \int_0^\infty
x^{t-1} e^{-x} dx$ the gamma function. \et \bpr Let
\[
 C_i(z_i,\xi_i) = z_i^{\xi_i} \frac{\Gamma \left(\frac{m}{2}\right)}{2^{\xi_i} \Gamma
\left(\frac{m}{2}+\xi_i\right)}
\]
One verifies easily
that
\[
 K_i^{a, (m)} C_i = C_i \caK_i^{a,(m)}
\]
for $a= +, -,0$. The proof then continues as in section
\ref{dualex}. \epr
\subsection{The instantaneous thermalization limit and the KMP process}
In the KMP model, introduced in \cite{KMP}, the energies $E_i$ of
different sites $i\in {\cal S}$ are updated by selecting a pair of
lattice sites $(i,j)$ and uniformly redistributing the energy under
the constraint of conserving $E_i + E_j$. In this section we show
that the KMP model arises by taking what we call here an
instantaneous thermalization limit of the process with generator
$L^{(m)}$, {\em for the case $m=2$}.

We start by computing the stationary measure of the process with
generator $L^{(m)}_{ij}$. \bl\label{statlem} Let $(z_i (t),z_j(t))$
be the Markov process with generator $L^{(m)}_{ij}$, starting from
an initial condition $(z_i (0), z_j (0))$ with $z_i (0)+ z_j (0)=E$.
Then in the limit $t\to\infty$ the distribution of $(z_i
(t),z_j(t))$ converges to the distribution of the couple
$((E+\epsi)/2, (E-\epsi)/2)$ where $\epsi$ has probability density
\be\label{statdens} f(\epsi) = C_m (E^2-\epsi^2)^{\frac{m}2 -1} \ee
$-E\leq \epsi\leq E$ and $f=0$ otherwise, and where $C_m$ is the
normalizing constant. \el \bpr Define $(E(t), \epsi (t)) = ( z_i (t)
+ z_j (t) , z_i (t) -z_j (t))$. Then simple rewriting of
$L^{(m)}_{ij}$ in the new variables yields that $(E(t), \epsi(t))$
is a Markov process with generator \be\label{croki} \loc'=
4(E^2-\epsi^2)\partial_\epsi^2 - 4m\epsi \partial_\epsi \ee From the
form of $\loc'$ we see immediately that $E$ is conserved and that
for given $E$, $\epsi (t)$ is an ergodic diffusion process with
stationary measure solving \be\label{kolmo}
\partial_\epsi^2 ( 4(E^2-\epsi^2) f) + \partial_\epsi (4m\epsi f) =0
\ee
Now notice that the rhs of
\eqref{statdens} solves
\[
 \partial_\epsi ( 4(E^2-\epsi^2) f) +  (4m\epsi f) =0
\]
and hence \eqref{kolmo}. \epr Denote by $\gamma_m$ the distribution
of $((E+\epsi)/2, (E-\epsi)/2)$. We can now define what me mean by
instantaneous thermalization. \bd Let $f: [0,\infty)^{{\cal S}}\to
\R$. For $e=(e_i)_{i\in {\cal S}}$ a configuration of energies,
$(i,j)\in {\cal S}\times {\cal S}$,
$(e'_i,e'_j)\in[0,\infty)\times[0,\infty)$ we denote by $
t(e,e'_i,e'_j)$ the configuration obtained from $e$ by replacing
$e_i$ by $e'_i$ and $e_j$ by $e'_j$. The instantaneous
thermalization of a pair $(i,j)\in {\cal S}\times {\cal S}$ is
defined by \be \caT^{(m)}_{ij} f(e) = \int f(t(e,e'_i,e'_j))
d\gamma_ m(e'_i,e'_j) \ee \ed

The instantaneously thermalized version of the Brownian energy
process is then defined as the process with generator
\be\label{kmpgen} \loc^{IT}_m f(e) = \sum_{ij\in {\cal S}} p(i,j)(
\caT^{(m)}_{ij} f(e) - f(e)) \ee This means that with rate $p(ij)$ a
pair $(i,j) \in {\cal S}\times {\cal S}$ is chosen and the energy is
instantaneously thermalized according to the measure $\gamma_m$.
From \eqref{statdens} one sees that, for $m=2$, the uniform
redistribution of the KMP model is recovered.

It is interesting to consider the dual of the instantaneous
thermalization process for general $m\in\N$. From the previous
discussion one knows that in the case $m=2$ this is just the dual of
the KMP model. However, the model with generator has for general $m$
a dual with different duality functions as is shown in Theorem
\ref{kmpthm} below. To introduce this dual, we remind the reader
that the Brownian energy process with generator $L^{(m)}$ is dual to
the discrete particle jump process with generator $L_{dual}^{(m)}$.
The following lemma gives the stationary measure of the dual BEP,
which is needed in the construction of the instantaneous thermalized
version of the dual BEP. \bl\label{statdualenij} Let $(k_t, l_t)$
evolve according to the generator $\loc_{dual}^{ij}$, and suppose
that initially $ k_0 +l_0 =N$, then in the limit $t\to\infty$,
$(k_t, l_t)$ converges in distribution to $((N+\Delta)/2,
(N-\Delta)/2)$ where $\Delta$ has distribution $\mu$ on $\{-N, -N+2,
\ldots, N\}$ with \be\label{recmu} \frac{\mu (\Delta)}{\mu
(\Delta-2)} = \frac{(N-\Delta +1) (N+\Delta -1 +m)}{(N+\Delta)
(N-\Delta + m)} \ee In particular for $m=2$, $(k_t, l_t)$ converges
to the uniform measure on the set $\{(a,b)\in\{ 0,\ldots, N\}: a+b =
N\}$. \el \bpr The process $(N_t, \Delta_t ) := ( k_t + l_t,
k_t-l_t)$ performs transitions $(N,\Delta) \to (N,\Delta-2)$ at rate
$ \frac14( N+\Delta) (N-\Delta +m)$ and $(N,\Delta) \to
(N,\Delta+2)$ at rate $\frac14 (N-\Delta) (N+\Delta + m)$. The
marginal $\Delta_t$ is then an irreducible continuous-time Markov
chain on the set $\{-N, -N+2, \ldots, N\}$, and hence has a unique
stationary measure. Since it is a pure birth and death chain, this
measure is also reversible. The recursion \eqref{recmu} then follows
from detailed balance. \epr

We denote by $\hat{\gamma}_m (k,l)$ the stationary distribution of
lemma \eqref{statdualenij}. For $\xi\in \N^{\cal S}$, and $(i,j)\in
{\cal S}\times{\cal S}$, $(\xi'_i,\xi'_j)\in \N\times\N$ we denote
by $t(\xi,\xi'_i,\xi'_j)$ the configuration obtained from $\xi$ by
replacing the value at $i$ by $\xi'_i$ and at $j$ by $\xi'_j$. We
then define the dual thermalization by \be\label{dualtherm}
\caT^{dual,(m)}_{ij} f(\xi) =\sum_{\xi'_i,\xi'_j: \;\;\xi'_i+\xi_j'
= \xi_i + \xi_j} f(t(\xi,\xi'_i,\xi'_j)) \hat{\gamma}_m
(\xi'_i,\xi'_j) \ee and the dual instantaneously thermalized energy
process as the process with generator \be \loc^{IT,(m)}_{dual} f
(\xi) = \sum_{ij\in {\cal S}} (\caT^{dual,(m)}_{ij} f(\xi) - f(\xi))
\ee

\bt\label{kmpthm} Consider the instantaneously thermalized version
of the Brownian energy process, with generator $\loc^{IT}_m$. This
process is dual to the process with generator $\loc^{IT,
(m)}_{dual}$ with duality function given by \be\label{marcuze}
D(e,\xi) = \prod_i D_i(e_i,\xi_i) \ee \be\label{poe} D_i(e_i,\xi_i)=
e_i^{\xi_i} \frac{\Gamma (m/2)}{2^{\xi_i} \Gamma (m/2+\xi_i)} \ee
\et \bpr By the duality result for the Brownian energy process,
Theorem \ref{ult}, we have for all $(i,j)\in {\cal S}\times{\cal S}$
\be L^{(m)}_{ij} D(e,\xi ) = L^{(m),dual}_{ij} D(e,\xi) \ee
therefore, \be \lim_{t\to\infty}(e^{tL^{(m)}_{ij}} -id) D(e,\xi ) =
\lim_{t\to\infty} (e^{tL^{(m),dual}_{ij}}-id) D(e,\xi)\;. \ee The
result then follows from the definition of the processes, together
with lemma \ref{statlem} and lemma \ref{statdualenij}. \epr
\subsection{Limiting processes as $m\to\infty$}
\label{kinf} As it was done for the $2j$-SEP, we study here the
limiting behavior of the $m$-BEP process for large $m$. \bt Consider
the process $\{z_t^{(m)}: t\ge 0\}$ with generator $L^{(m)}$ and
initial condition $z^{(m)} \in \R_+^{{\cal S}}$ and its dual
$\{\xi_t^{(m)}: t\ge 0\}$ with generator $L_{dual}^{(m)}$ and
initial condition $\xi^{(m)}\in\N^{{\cal S}}$. Suppose that, as
$m\to \infty$, $z^{(m)} \to z \in \R_+^{{\cal S}}$ and
$\xi^{(m)}\to\xi \in\N^{{\cal S}}$. Then: \begin{enumerate} \item
the process $\{z_{t/m}^{(m)}: t\ge 0\}$ converges to the process,
$(z_{t})_{t\ge 0}$ started from $z$, with generator
$$
L = \sum_{ij\in {\cal S}} p(i,j) L_{ij}
$$
$$
L_{ij} = - 2 (z_i-z_{j}) \left(\frac{\partial}{\partial z_i} -
\frac{\partial}{
\partial z_{j}}\right)
$$
\item the process $\{\xi_{t/m}^{(m)}\}$ converges to a system of independent
random walkers $(\xi_{t})_{t\ge 0}$ started from $\xi$, with
generator
$$
L_{dual} = \sum_{ij\in {\cal S}} p(i,j) L_{ij}^{dual}
$$
$$
L_{ij}^{dual} = 2\xi_i (f(\xi^{ij})-f(\xi)) +
2\xi_j(f(\xi^{ji})-f(\xi))
$$
\item
The two limiting processes $(x_t)_{t\ge 0}$ and $(\xi_t)_{t\ge 0}$
above are each other dual, with duality function $$ D(x,\xi) =
\prod_{i\in {\cal S}} x_{i}^{\xi_i}$$ \end{enumerate}\et \bpr The
proof of items 1. and 2. proceeds like in Theorem \ref{jinf}. For
item 3. compare to example in section \ref{dualdet}.  \epr

\subsection{Boundary driven process}

In this last section we consider the $m$-BEP process in contact at
its boundary to energy reservoirs of the Ornstein-Uhlenbeck type. A
duality result for the Brownian Momentum Process with reservoirs was
already proven in \cite{gkr}. Here we generalize this result to the
general Brownian energy process for arbitrary $m\in\N$. We start
from the momentum process $\{(x(t)_{i,\alpha}): i\in {\cal S},
\alpha=1,\ldots,m, t\ge 0\}$ on a ladder graph with $m$ levels and
all levels at sites $i\in{\partial {\cal S}}$ connected to a
thermalizing Ornstein-Uhlenbeck process which parameter $T_i$, to be
thought as the temperature. The generator reads
\begin{eqnarray}\label{bdladder} L & = & \sum_{i,j\in {\cal S}}
p(i,j) \sum_{\alpha,\beta=1}^m
\left(x_{i,\alpha}\frac{\partial}{\partial x_{j,\beta}} -
x_{j,\beta}\frac{\partial}{\partial x_{i,\alpha}}\right)^2 \nonumber \\
& + & \sum_{i\in\partial{\cal S}}\sum_{\alpha=1}^m T_i
\frac{\partial^2}{\partial x_{i,\alpha}^2}  -
x_{i,\alpha}\frac{\partial}{\partial x_{i,\alpha}}
\end{eqnarray}
If we now consider the induced process $\{z_i(t): i\in {\cal S},
t\ge 0\}$ measuring the energy at each site via the map
$$
z_i = \sum_{\alpha=1}^m x_{i,\alpha}^2\;,
$$
then, using the identities \eqref{ident}, we find the generator
\begin{eqnarray}\label{bdm} L & = & \sum_{i,j\in {\cal S}} p(i,j) 4 z_i
z_j\left(\frac{\partial}{\partial z_i} -\frac{\partial}{\partial
z_{j}}\right)^2 - 2m (z_i-z_{j}) \left(\frac{\partial}{\partial z_i}
- \frac{\partial}{\partial z_{j}}\right)\nonumber \\
& + & \sum_{i\in\partial{\cal S}} 2
T_i\left(m\frac{\partial}{\partial z_i} + 2
z_i\frac{\partial^2}{\partial z_i^2} \right) -
2z_i\frac{\partial}{\partial z_i}
\end{eqnarray}
Introducing as usual a set $\partial_e {\cal S}$ of sink sites and a
bijection $i\mapsto i_e$ which associate each boundary site $i\in
\partial {\cal S}$ to a sink site $i_e\in\partial_e {\cal S}$,
we have the following duality theorem: \bt\label{KMPboundthm}
Let $(z_t)_{t\ge 0}$ denote the boundary driven $m$-BEP with
generator \eqref{bdm}. Then $(z_t)_{t\ge 0}$ is dual to the process
$(\xi_t)_{t\ge 0}$ with generator
\begin{eqnarray}\label{bdmdual} L_{dual} f(\xi) & = & \sum_{i,j\in {\cal S}} p(i,j)
2\xi_i (2\xi_j+m) (f(\xi^{i,j})-f(\xi)) + 2\xi_j (2\xi_i+m)
(f(\xi^{j,i})-f(\xi)) \nonumber \\
& + & \sum_{i\in\partial {\cal S}} 2 \xi_i
(f(\xi^{i,i_e})-f(\xi))\end{eqnarray} with duality function given by
\be\label{bdmdualfct} D(z,\xi) = \prod_{i\in\partial_e {\cal S}}
T_i^{\xi_i}\prod_{i\in {\cal S}}z_i^{\xi_i} \frac{\Gamma
\left(\frac{m}{2}\right)}{2^{\xi_i} \Gamma
\left(\frac{m}{2}+\xi_i\right)}  \ee \et \bpr The bulk part of the
duality function coincides with the one of Theorem \ref{ult}; the
boundary part is easily checked with an explicit computation. \epr

\end{document}